\documentstyle[prl,aps,twocolumn]{revtex}

\def\be{\begin{equation}}
\def\ee{\end{equation}}
\def\bex{\begin{eqnarray}}
\def\eex{\end{eqnarray}}

\def\hi{{\cal H}}

\begin{document}
\draft

\title{Binding entanglement channels}

\author{Pawe\l{} Horodecki$^1$ \cite{poczta1}, Micha\l{} Horodecki$^2$,
\cite{poczta2} Ryszard Horodecki$^2$\cite{poczta3} }

\address{$^1$ Faculty of Applied Physics and Mathematics,
Technical University of Gda\'nsk, 80--952 Gda\'nsk, Poland\\
$^2$ Institute of Theoretical Physics and Astrophysics,
University of Gda\'nsk, 80--952 Gda\'nsk, Poland}

\maketitle

\begin{abstract}
We  define the binding entanglement channel as the quantum channel
through which quantum information cannot be reliably transmitted, but
which can be used  to share bound entanglement. We provide a characterization of
such class of channels. We also show that any bound entangled state can
be used to construction of the map corresponding the binding
entanglement channel.
\end{abstract}

\section{Introduction}

One of the recent results leading to better understanding of
quantum entanglement \cite{EPR,Schrodinger} was realizing that there are two
qualitatively different types of entanglement of mixed states of two-component
systems \cite{bound,Pawel}.
Namely,
there is {\it free entanglement} (FE) which can be converted into pure singlet
form by means of local quantum operations and classical communication (LQCC).
Such a process is called distillation \cite{Bennett_pur} and it allows to use
the noisy entanglement for the purposes of quantum communication. However,
there is also {\it bound entanglement} (BE), which cannot be distilled
\cite{bound,Pawel}.
At present the
structure and properties of BE state are being extensively investigated
\cite{aktyw,Popescu99,upb,Terhal,single,Rains}. In particular, a
striking connection between the bound etanglement and nonlocality without
entanglement \cite{nonlocality} has been discovered \cite{upb}.
Also, the bound
entanglement implies
a new approach in entanglement measures: one must, in general  leave the
paradigm that a measure of entanglement should vanish only on separable
states. Indeed, at present we know that physically the most relevant
measure of entanglement \cite{huge,Rohrlich,Plenio} which is
distillable entanglement
does not satisfy this condition (it vanishes on the bound entangled states).
The above,  more general approach allowed to obtain a new bound on 
distillable entanglement
 Ref. \cite{Rains}. Due to the connection between entanglement
and positive maps \cite{sep} the investigation of bound entanglement
was also fruitful for pure mathematics. Namely, by use of results
on bound entanglement of Ref. \cite{upb} the  first {\it systematic} way
of constructing the so called non-decomposable positive maps was
found in Ref. \cite{Terhal}.

In this paper we would like to investigate the {\it processes} of
interaction with environment, which lead to bound entanglement.
In general, the mixed states emerge from
interaction with environment, which is very hard to be avoided in realistic
situation. Such interaction may completely destroy the initial pure
entanglement, or sometimes there may remain some residual entanglement, free
or bound. We will be interested in the processes for which the residual
entanglemet is the bound one.
To be more
precise, imagine that Alice can send particles to Bob via a quantum channel
$\Lambda$ (representing the interaction with environment). Alice and Bob
are allowed to support the quantum channel by using LQCC operations 
and can
enhance the transmission by sending entangled particles down the channel.
The latter means that  effectively they have a channel
$\Lambda^{\otimes N}$ for arbitrary $N$.
Now we are interested in such channels that Alice and Bob
(i) cannot send reliably quantum information (equivalently, cannot produce
asymptotically singlet state);
(ii) can produce a BE state. Such channels we will call {\it binding
entanglement channels} (BE channels).

We prove a theorem characterizing such channels, which says that a channel
is BE if and only if sending half of maximally entangled pair through
the channel, one obtains BE state.
It follows that a channel is
BE if there exists a  pure  entangled  state such, that if sent through the
channel it becomes bound entangled. Thus knowing the examples of 
BE states, we can
construct the BE channels.  We provide a way of constructing BE channel from
any given BE state. Our investigations are based on the
general connections between channels and bipartite states investigated in
\cite{Jamiol,huge,Nielsen,xor,single}.

\section{Binding entanglement channels: characterization}

To begin with, let us introduce some notation. By a channel we mean any
completely positive (CP) trace-preserving map.
A completely positive  map  $\Lambda:M_m\rightarrow M_n$ will be denoted by
$\Lambda_m^n$ (here $M_n$ denotes the set of $n\times n$ square matrices.
The identity map acting on $M_n$ will be denoted by $I_n$.
 Maximally entangled state on the system $M_n\otimes M_n$ of the form
\be
P_+^n={1\over \sqrt n}\sum_{i=1}^n |i\rangle|i\rangle
\ee
will be called  singlet state. A state acting on
the Hilbert space $C^m\otimes C^n$ will be denoted
by $\varrho_{m,n}$ (or $\sigma_{m,n}$ etc.). Sometimes, if it does not
lead to misunderstanding  we will not write the indices explicitly.
Finally, $\varrho_{ikjl}$ denotes matrix element of the state $\varrho$ 
in product basis
\[
\varrho_{ikjl}\equiv \langle e_i\otimes
f_k|\varrho|e_j\otimes f_l\rangle.
\]

{\bf Definition.} \it We say that a channel $\Lambda$ is \rm binding
entanglement channel \it iff (i) $Q_2(\Lambda)=0$ and (ii) it is possible to
obtain bipartite bound entangled state by means of (possibly multiply) use of
the channel and LQCC operations. \rm

Here $Q_2$ is the quantum capacity of a channel supported by LQCC action (the
subscript $2$ indicates two-way classical communication) \cite{huge}.
Now we will prove a theorem characterizing such channels in terms of
BE bipartite states.

{\bf Theorem.} A channel $\Lambda:M_m\rightarrow M_n$ is binding entanglement
iff the state $(I\otimes \Lambda)P_+^m$ (acting on $C^m\otimes C^n$)
is a BE state.

{\bf Proof.} Let us first prove the  sufficiency of the condition. If
$(I\otimes \Lambda_m^n)P_+^m$ is BE state then (ii) is obviously satisfied, so
that one needs to prove that the condition implies also (i). Suppose,
conversely, that $Q_2(\Lambda_m^n)>0$. Then, one can produce asymptotically pure
singlets by use of the channel and LQCC. The first stage of the most
general protocol of producing singlet pairs is sending half of some
state $\sigma_{{k\times N},{m\times N}}$ via
the channel $\Lambda^{\otimes N}$ (denote it by
$\Lambda_{m\times N}^{n\times N}$). The second stage
amounts to distillation of the emerging state
$\varrho_{{k\times N},{m\times N}}=(I_{k\times N}\otimes
\Lambda_{m\times N}^{n\times N}) \sigma$.
Hence, to obtain finally the singlets, the state $\varrho$  must be FE.
We will now show that this implies that $(I_m\otimes \Lambda_m^n) P_+^m$
must be also FE. To see it, note that the   state $\sigma$ (as any state)
can be written as
$\sigma=(\Gamma_{m\times N}^{k\times N}\otimes I_{m\times N})P_+^{m\times N}$
(where $\Gamma$ is CP, but not necessarily trace-preserving map).
So we have
\bex
&&\varrho=(I_{k\times N}\otimes \Lambda_{m\times N}^{n\times N})
(\Gamma_{m\times N}^{k\times N}\otimes I_{m\times N}) P_+^{m\times N}=
\nonumber\\
&&(\Gamma_{m\times N}^{k\times N}\otimes \Lambda_{m\times N}^{n\times N})
P_+^{m\times N}=\nonumber\\
&&(\Gamma_{m\times N}^{k\times N}\otimes I_{n\times N})(I_{m\times N}\otimes
\Lambda_{m\times N}^{n\times N})P_+^{m\times N}
\eex
Now, since $\varrho$ is FE, then also $(I_{m\times N} \otimes
\Lambda_{m\times N}^{n\times N})P_+^{m\times N}$ must be
FE (indeed the action $\Gamma\otimes I$ is LQCC one, hence cannot produce FE
state from a BE one). Now, since $(I_{m\times N}\otimes
\Lambda_{m\times N}^{n\times N})P_+^{m\times N}=
\left( (I_m\otimes \Lambda_m^n)P_+^m\right)^{\otimes N}$ we obtain that
also $(I_m\otimes\Lambda_m^n)P_+^m$ must be FE, which is a contradiction.
Hence, if $(I_m\otimes\Lambda_m^n)P_+^m$ is BE then the condition (i)
is satisfied.

Now, we will show that the condition that
$(I_m\otimes\Lambda_m^n)P_+^m$ is BE is also a necessary one for $\Lambda$
to be BE.  Suppose, conversely,  that $(I_m\otimes\Lambda_m^n)P_+^m$ is not BE.
Then it can be separable or FE. If its is FE, then one can distill it and
obtain nonzero $Q_2$ so that the condition (i) is violated. If, instead
$(\Lambda_m^n\otimes I_m)P_+^m$ is separable, then we will show that the
condition (ii) is violated. Indeed, if for some state $\sigma_{k,m}$ the state
$\varrho_{m,n}=(I_m\otimes\Lambda_m^n)\sigma_{k,m}$ is BE, then
writing $\sigma$ as $\sigma_{k,m}=(\Gamma_k^m\otimes I_m)P_+^m$ we
obtain, similarly as in the proof of sufficiency, that
$\varrho=(\Gamma_k^m\otimes I_n)(I_m\otimes\Lambda_m^n)P_+^m$.
  Then, since $\Gamma\otimes I$ is LQCC, we obtain that
$(I_m\otimes\Lambda_m^n)P_+^m$ cannot be separable (LQCC action cannot make BE
state from separable one). This ends the proof.

From the above characterization of BE channels  it follows that
given a channel with  $Q_2=0$, if bound entanglement can be created at
all, then it can be created without exchange of classical information between
Alice and Bob but merely by sending half of singlet pair through the channel.
Hence also multiply use of channel is not needed.

\section{Binindg entanglement channels from bound entangled states}
In this section we will provide a procedure of constructing BE channels
from BE states.
As one knows there is an
isomorphism between the set of states $\varrho_{m,n}$ with maximally mixed
reduction $\varrho_A$ and the channels $\Lambda_m^n$.
It is given just by the formula:
\be
\varrho_{m,n}=(I_m\otimes \Lambda_m^n)P_+^m
\label{isom1}
\ee
(the maximally mixed reduction is connected with the fact that channels
preserve trace).
In other words, if one has a channel, one can send half of singlets through it
to obtain the state with maximally mixed reduction, and, conversely,
any state of maximally mixed reduction emerges from sending half of singlet
down some channel. Explicitly, the connection between matrix elements of state
and associated channel is the following
\be
\langle f_k|\Lambda(|e_i\rangle\langle
e_j)|f_l\rangle\equiv\lambda_{klij}=
\varrho_{ikjl}\equiv \langle e_i\otimes
f_k|\varrho|e_j\otimes f_l\rangle.
\label{isom}
\ee
So we can provide examples of BE channels basing on the known BE states
with maximally mixed reduction. However, one also knows the examples of BE
states with none of reductions maximally mixed \cite{upb}. How to associate
channels with them? As mentioned above, the maximally mixed reduction
is connected with the fact that
the channel acts only on one half of the singlet, so that,
being trace-preserving, it cannot disturb the other one. Since the singlet is
maximally entangled, it has  maximally mixed reduction that is inherited by
the final state. Now, if a state with {\it non-maximally}
mixed reductions is concerned, one can imagine it emerges from sending
{\it non-maximally entangled pure state} via a channel. The state must have
the same reduction as the mixed state of interest (as, again, the channel will
not affect that reduction). To recover such a channel from the given state
$\varrho$, we will first transform it into a state $\sigma$ of maximally
mixed reduction by means of LQCC action. Then the channel will be the one
associated with $\sigma$ via the state-channel isomorphism (\ref{isom1}).
Let a BE state  $\varrho_{m,n}$ acts on $\hi_A\otimes \hi_B$ and
let $\hi'_A$ be the
support of its reduction $\varrho_A$ with $\dim\hi'_A=k$. Then define
\be
\sigma_{r,n}=(r\varrho_A)^{-1/2}\otimes I\,\varrho\,(r\varrho_A)^{-1/2} 
\otimes I,
\ee
where $\varrho_A$ was inverted on its support $\hi_A'$
Here we used the fact that the support of any state is equal to
product of the supports of its reductions (see Appendix), so that,  in
 fact, both
$\varrho$ and
$\sigma$ acts on $\hi'_A\otimes\hi_B$. It is easy to check
that $\sigma_A={I\over r}$. Indeed,
choosing the basis $\{e_i\}\subset\hi_A^i$ to be eigenbasis of 
$\varrho_A$ (i.e.
$\varrho_A=\sum_ip_i|e_i\rangle\langle e_i|$) we obtain
\be
\sigma_{ikjl}={1\over r\sqrt {p_ip_j}}\varrho_{ikjl},
\ee
hence
\be
(\sigma_A)_{ij}=\sum_k\sigma_{ikjk}={1\over r\sqrt {p_ip_j}} p_i\delta_{ij}=
{1\over r} \delta_{ij}
\ee
Now, as the state $\sigma$ was created from $\varrho$ by LQCC action, then
it is BE (the action is called filtering \cite{conc}). Then the
seeked BE channel $\Lambda_A$ corresponding to the given state $\varrho$ is the
one associated with the state $\sigma$ via the formula (\ref{isom})
(the subscript $A$ indicates that we recover the channel  by use of the reduction
$\varrho_A$).

Then to obtain explicit form of $\Lambda_A$ one needs to calculate the
map $\Theta$ given by the formula
\be
(I_r\otimes \Theta_r^n)P_+^r=\varrho.
\label{map-state}
\ee
Then $\Lambda_A$ is given by
\be
\Lambda_A= \Theta\circ \Gamma^T_A
\ee
where
$\Gamma_A(\cdot)={1\over r} \varrho_A^{-1/2}
(\cdot)\varrho_A^{-1/2}$ and $T$ is tranpose in the space of
maps i.e. $(\Theta_B^T)_{klij}=(\Theta_B)_{ijkl}$. If the given map $\Lambda$
is CP (as in our case) and its Stinespring form is
$\Lambda(\cdot)=\sum_iV_i(\cdot)V_i^\dagger$ then the transposed map is
given simply by $\Lambda(\cdot)=\sum_iV^T_i(\cdot){(V_i^T)}^\dagger$.
Thus we obtain that
\be
\Lambda_A(\cdot)={1\over
r}\Theta((\varrho_A^T)^{-1/2}(\cdot)(\varrho_A^T)^{-1/2}).
\label{odzysk}
\ee
Note that if only $\varrho_A$ is not maximally mixed then both $\Gamma$ and
$\Theta$ are not trace-preserving. Nevertheless $\Lambda$ is trace-preserving
so that it constitutes a channel.

Of course one can use the other reduction of the state $\varrho$ to obtain a
channel (call it $\Lambda_B$). Then one can get  the following formula
\be
\Lambda_B(\cdot)={1\over r}
\Theta^T((\varrho_B^T)^{-1/2}(\cdot) (\varrho_B^T)^{-1/2}).
\ee
Now, the state $\varrho$ emerges if (i) Alice send to Bob some  pure state $\psi$
of both reductions equal to $\varrho_A$ via the channel $\Lambda_A$, or (ii)
Bob sends to Alice the pure state of reductions $\varrho_B$
through the channel $\Lambda_B$.

\section{Examples}
A simple way of recovering the maps from a given state
via formula  (\ref{map-state}) is to use the eigenbasis of the
state. Namely, if
\be
\varrho_{m,n}=\sum_ip_i|\psi_i\rangle\langle \psi_i|
\ee
with 
\be
\psi_i=\sum_{j=1}^m\sum_{k=1}^n c^i_{j,k}|e_i\rangle|f_i\rangle,
\ee
then the associated map $\Theta_m^n$ is given by
\be
\Theta(\cdot)=\sum_ip_iV_i(\cdot)V_i^\dagger
\ee
with $\langle e_j|V_i|f_k\rangle=mc^i_{j,k}$. If it is hard to find the
eigenbasis, one can use any decomposition of the BE state into pure ones.

{\it Example 1.}
In the paper Ref. \cite{bound} we modified the St\o{}rmer \cite{Stormer}
matrix  to obtain the following family of two-qutrit BE states
\be
\sigma_{\alpha}=\frac{2}{7}P_+^3
+\frac{\alpha}{7} \sigma_{+}+
\frac{5-\alpha}{7} \sigma_{-},
\label{target}
 \ee
where $3<\alpha\leq4$ and
\begin{eqnarray}
\sigma_{+}={1 \over 3}(|0\rangle|1\rangle \langle 0| \langle 1|
+ |1\rangle|2\rangle \langle 1| \langle 2|+
|2\rangle|0\rangle \langle 2| \langle 0|), \nonumber \\
\sigma_{-}={1 \over 3}(|1\rangle|0\rangle \langle 1| \langle 0|
+ |2\rangle|1\rangle \langle 2| \langle 1|+
|0\rangle|2\rangle \langle 0| \langle 2|).
\label{mix}
\end{eqnarray}
The above state has both reductions maximally mixed, so that we could consider
two channels ($\varrho=(I\otimes \Lambda_1)P_+$ and
$\varrho=(\Lambda_2\otimes I)P_+$). However, due to symmetry of the state, the
two cases give raise to the same family of channels, given by
\bex
&&\Lambda(\cdot)= {2\over 7} (\cdot)+ {\alpha\over 7} \sum_{k=1}^3 P_{k\oplus 1
k}(\cdot)P_{kk\oplus1}+\nonumber\\
&&{5-\alpha\over 7} \sum_{k=1}^3 P_{k\ominus1
k}(\cdot)P_{kk\ominus1}
\eex
where $P_{ij}=|i\rangle\langle j|$; $\oplus$ and $\ominus$ denote + and $-$
modulo 3 respectively.

{\it Example 2.}
This example will be based on the two-qutrit BE state \cite{Pawel} of the
following form
\be
\varrho={1 \over 8a + 1}
\left[ \begin{array}{ccccccccc}
	  a &0&0&0&a&0&0&0& a	\\
	   0&a&0&0&0&0&0&0&0	 \\
	   0&0&a&0&0&0&0&0&0	 \\
	   0&0&0&a&0&0&0&0&0	 \\
	  a &0&0&0&a&0&0&0& a	  \\
	   0&0&0&0&0&a&0&0&0	 \\
	   0&0&0&0&0&0&{1+a \over 2}&0&{\sqrt{1-a^2} \over 2}\\
	   0&0&0&0&0&0&0&a&0	 \\
	  a &0&0&0&a&0&{\sqrt{1-a^2} \over 2}&0&{1+a \over 2}\\
       \end{array}
      \right ], \ \ \
\ee
where $0<a<1$. The reduction $\varrho_A$ of the state is given by
\be
\varrho_A={1\over 8a+1}\left[\begin{array}{ccc}
3a&0&0\\
0&3a&0\\
0&0&{1+2a}
\end{array}\right]
\ee
hence it is not maximally mixed.
Now, to recover the channel we can apply the formula (\ref{odzysk}).
The map $\Gamma$ is given by
\bex
&&\Gamma(\cdot)= {a\over 8a+1} \left(3(\cdot) + P_{12} (\cdot) P_{21} +
P_{13} (\cdot) P_{31} \phantom{1\over a}
+ P_{21} (\cdot) P_{12}+\right.\nonumber\\
&& \left. \phantom{1\over a}
P_{23} (\cdot) P_{32}
+ P_{32} (\cdot) P_{23}\right) + {1\over 8a+1} W (\cdot) W^\dagger
\eex
where  $W=\sqrt{1+a\over2}P_{31}+
\sqrt{1-a\over 2}P_{33}$.
Since $\varrho_A$ is diagonal we obtain $\Gamma_A^T=\Gamma_A$
Then the final form of the channel $\Lambda_A$ is given by
\bex
&&\Lambda_A (\cdot)={a\over3} \left(3V(\cdot)V +
{1\over 3a}(P_{12} (\cdot) P_{21} +
P_{32} (\cdot) P_{23} \phantom{1\over a} \right.\nonumber\\
&&+ P_{21} (\cdot) P_{12})+ \left. {1\over 2a+1}(P_{13} (\cdot) P_{31}
+ P_{23} (\cdot) P_{32})\right) +  \tilde W (\cdot) \tilde W^\dagger
\eex
where
$V=diag[1/\sqrt{3a}, 1/\sqrt{3a}, 1/\sqrt{2a+1}]$
and $\tilde W=\sqrt{1+a\over6a}P_{31}+\sqrt{1-a\over 2(2a+1)}P_{33}$.

\section{Discussion}
Let us now discuss some possible directions of further
investigation of the binding entnaglement channels. The main goal will be
to find how the BE channels could be useful for quantum communication. The
hint is given by the effect of activation of bound entanglement \cite{aktyw},
where a large amount of BE systems considerably raised the possibilities of
a single FE system. In Ref. \cite{aktyw} we rose a question, whether
the channels associated with the BE states (which, due to theorem, are BE
channels) could exhibit nonadditivity in the following sense. If we have a
channel of some nonzero capacity $Q$, and a BE channel, then by using the
channels jointly, one expects to obtain total capacity greater than $Q$.

Another question arises, if we consider the BE channel as public one (cf.
\cite{aktyw}). This changes the paradigm of entanglement manipulations,
where so far, only classical communication was public. The question is:
what is capacity of some quantum channel of nonzero standard capacity (either
with or without classical comunication) if supplemented with public
BE channel?
It was natural  to expect  that the capacity of the supported
channel  could  be  strictly greater, especially, because,
as reported in Ref. \cite{upb}, the BE states can have
surprisingly large entanglement of formation ($E_f$). The two-qutrit states
provided in Ref. \cite{upb} have $E_f\simeq 0.2$ of
entanglement of formation  while the maximally entangled state of
two-qutrits has $E_f\simeq 1.5$. Now we would like to ask the following
question: may be, {\it the channel supported by public BE channel have
maximal capacity, determined only by the Hilbert space of the sent systems}?
This could be called the effect of cristallization of bound entanglement.
The conjecture is not unreasonable: we have, in fact, infinite amount of
bound entanglement at our disposal.

\begin{appendix}
\vskip5mm
\centerline{\large Appendix}
\vskip2mm
Here we prove the following lemma:

{\bf Lemma.} The support of any state is included in the product of the
supports of the reductions of the state
\be
{\rm supp} \varrho\subseteq {\rm supp}
\varrho_A\otimes {\rm supp} \varrho_B
\ee
{\bf Proof.}
Let us first prove the lemma for pure state $|\psi\rangle\langle\psi|$.
Writing the state in the Schmidt decomposition we see that it is a
superposition of the products of the states belonging to  supports of
the reductions, so that the thesis of the lemma holds. Now, for
the mixed state
$\varrho=\sum_i|\psi_i\rangle\langle\psi_i|$
we have
\be
{\rm supp} \varrho={\rm span} \{\psi_i\}_i
\ee
and
\be
{\rm supp} \varrho_A \otimes {\rm supp} \varrho_B=
{\rm span} \{{\rm supp} \varrho^i_A \otimes {\rm supp} \varrho_B^i\}_i,
\ee
where $\varrho_{A,B}^i$ are the reductions of the states $\psi_i$.
Hence we obtain  the required inclusion.
\end{appendix}

\end{document}